\documentclass[nohyper,12pt,letterpaper]{JHEP3}

\usepackage{amsmath}
\usepackage{amsfonts}
\usepackage{amssymb}
\usepackage{graphicx}
\usepackage{epsfig}

\def\ssM{{\scriptscriptstyle M}}
\def\ssN{{\scriptscriptstyle N}}
\def\ssP{{\scriptscriptstyle P}}
\def\ssQ{{\scriptscriptstyle Q}}

\def\Box{{\hbox{$\sqcup$}\llap{\hbox{$\sqcap$}}}}

\def\be{\begin{equation}}
\def\ee{\end{equation}}
\def\bea{\begin{eqnarray}}
\def\eea{\end{eqnarray}}
\def\nn{\nonumber}
\def\ansatz{{\it ansatz}}

\def\exd{{\rm d}}
\def\pref#1{(\ref{#1})}
\def\endignore{}
\def\ignore #1\endignore{} % use to "comment out" text

\def\bd{\begin{displaymath}}
\def\ed{\end{diplaymath}}

\def\d{\mathrm{d}}

\def\cx{{\cal X}}
\def\cy{{\cal Y}}
\def\cz{{\cal Z}}

\def\ba{\begin{eqnarray}}
\def\ea{\end{eqnarray}}
\def\be{\begin{equation}}
\def\ee{\end{equation}}

\def\exd{{\rm d}}

\def\d{\mathrm{d}}
\def\({\left(}
\def\){\right)}

\title{Exact Wave Solutions to\\ 6D Gauged Chiral Supergravity}

\author{Andrew J. Tolley,${}^{1,{\rm a}}$
C.P. Burgess,${}^{1,2,3,{\rm b}}$ Claudia de Rham${}^{1,2,{\rm c}}$
and D. Hoover${}^{4,{\rm d}}$ \\
${}^1$ Perimeter Institute for Theoretical Physics, Waterloo ON,
N2L 2Y5, Canada.\\
${}^2$
Physics \& Astronomy, McMaster University, Hamilton ON, L8S 4M1,
Canada. \\
${}^3$ Theory Division, CERN, CH-1211 Geneva 23, Switzerland. \\
${}^4$ Physics Department, McGill University, Montr\'eal QC, H3A
2T8, Canada. \\ \vspace{-4mm} \\
e-mail: ${}^{{\rm a}}$\email{atolley@perimeterinstitute.ca}, ${}^{{\rm b}}$\email{cburgess@perimeterinstitute.ca}, ${}^{{\rm
c}}$\email{crham@perimeterinstitute.ca}, ${}^{{\rm d}}$\email{dhoover@perimeterinstitute.ca}}

\date{}
\abstract{We describe a broad class of time-dependent exact wave
solutions to 6D gauged chiral supergravity with two compact
dimensions. These 6D solutions are nontrivial warped
generalizations of 4D pp-waves and Kundt class solutions and
describe how a broad class of previously-static compactifications
from 6D to 4D (sourced by two 3-branes) respond to waves moving
along one of the uncompactified directions. Because our methods
are generally applicable to any higher dimensional supergravity
they are likely to be of use for finding the supergravity limit of
time-dependent solutions in string theory. The 6D solutions are
interesting in their own right, describing 6D shock waves induced
by high energy particles on the branes, and as descriptions of the
near-brane limit of the transient wavefront arising from a local
bubble-nucleation event on one of the branes, such as might occur
if a tension-changing phase transition were to occur.}

\preprint{PI-COSMO-65}

\begin{document}

\section{Introduction}

Understanding time-dependent dynamics is central to applications
of higher-dimensional supergravity theories \cite{HiDSugra} to
cosmology and to particle physics. Inasmuch as higher-dimensional
supergravities provide the low-energy limit of string theories,
any understanding of time-dependence in the supergravity limit
also provides a guide for the thornier issue of understanding
these same issues in string theory. For these reasons there is
considerable interest in finding time-dependent solutions to
higher-dimensional supergravity \cite{tdepsusy} (as well as of
non-supersymmetric gravity \cite{tdepnonsusy}, since this can also
sometimes capture similar physics).

Much of this activity has focussed on 10D and 5D theories
(motivated by string applications, and Randall-Sundrum \cite{RS}
constructions), although these can either be more difficult to
solve or they can have features which are specific to the relative
simplicity of co-dimension one spaces. Six-dimensional
supergravity has more recently emerged as being a useful
intermediate workshop within which to investigate phenomena which
can often generalize to still higher-dimensional contexts.
Interest in 6D supergravity has been further sharpened by the
recognition that it can provide insights into the nature of the
cosmological constant problem \cite{SLED1}--\cite{SLEDpheno}, by
building on the observation that higher-dimensional theories can
break the link between the 4D vacuum energy density and the
curvature of 4D spacetime
\cite{5DSelfTune}--\cite{6DNonSUSYSelfTunex} (see
\cite{Burgess:2007ui} for a review). There has also been
considerable recent interest in 6 dimensional models more
generally \cite{Closelyrelatedworks}. Including branes is
notoriously difficult due the necessity to regularize UV
divergences which arise \cite{regularizations}, however for recent
work on understanding this more deeply in the context of effective
field theory see \cite{claudiaeft}.

In this paper we further the program of understanding
time-dependent solutions to higher-dimensional supergravity in two
ways. In \S2, we present a method for constructing explicit exact
solutions to the supergravity field equations of 6D gauged chiral
supergravity, to derive a new class of exact solutions to these
equations which describe a class of gravitational waves passing
through spacetimes for which two dimensions are compactified in
response to the stress-energy of two space-filling 3-branes. In
\S3 we discuss the applications of these solutions, including how
to construct the shock wave metric corresponding to
ultra-relativistic particles and BH's moving on one of the branes,
and also how these solutions provide some insight into the
transient part of the dynamics describing outgoing waves which
would arise shortly after a phase transition on one of the source
branes.

\section{Wave Solutions to 6D Supergravity}

In this section we describe the technique of finding the wave solutions to the field equations of 6D gauged chiral supergravity \cite{NS,6DSugra,HiDSugra}.

\subsection{Field equations}

The action whose variation gives the field equations of interest
is  part of the Lagrangian density for 6D chiral gauged
supergravity, and is given by\footnote{The curvature conventions
used here are those of Weinberg's book \cite{GandC}, and differ
from those of MTW \cite{MTW} only by an overall sign in the
Riemann tensor.}
\bea \label{6DSugraAction}
    \frac{{\cal L}}{\sqrt{-g}} &=& - \frac{1}{2 \kappa^2} \,
    g^{\ssM\ssN} \Bigl[ R_{\ssM\ssN} + \partial_\ssM \phi \,
    \partial_\ssN \phi \Bigr] - \frac{2g^2}{\kappa^4} \; e^\phi \nn\\
    && \qquad\qquad - \frac14 \, e^{-\phi} \, F_{\ssM\ssN} F^{\ssM\ssN}
    - \frac{1}{2 \cdot 3!} \, e^{-2\phi} \, G_{\ssM\ssN\ssP}
    G^{\ssM\ssN\ssP} \,,
\eea
where $\phi$ is the 6D scalar dilaton, $G = \exd B$ is the field
strength for a Kalb-Ramond potential and $F = \exd A$ is the field
strength for the gauge potential, $A_\ssM$, whose flux in the
extra dimensions stabilizes the compactifications. The parameters
$g$ and $\kappa$ have dimensions of inverse mass and inverse
mass-squared, respectively. These expressions set some of the
bosonic fields of 6D supergravity to zero, as is consistent with
the corresponding field equations. The corresponding field
equations are $E = E_{\ssM} = E_{\ssM\ssN} = E_{\ssM\ssN}^a = 0$,
where:
\bea \label{fieldequations}
    &&E = \Box \, \phi + \frac{\kappa^2}{6} \, e^{2 \phi} \;
    G_{\ssM\ssN\ssP} G^{\ssM\ssN\ssP} +
    \frac{\kappa^2}{4} \, e^{-\phi} \; F_{\ssM\ssN} F^{\ssM\ssN}
    - \frac{2 \,g^{2}}{\kappa^2} \, e^{\phi} \nn \\
    &&E_a^{\ssN\ssP} = D_\ssM \Bigl(e^{ -2 \phi}
    \, G^{\ssM\ssN\ssP} \Bigr) \nn\\
    &&E^\ssN = D_\ssM \Bigl(e^{ - \phi} \, F^{\ssM\ssN} \Bigr) + e^{-2 \phi}
    \, G^{\ssM\ssN\ssP} F_{\ssM\ssP}  \\
    &&E_{\ssM\ssN} = R_{\ssM\ssN} + \partial_\ssM\phi \, \partial_\ssN\phi
    + \frac{\kappa^2}{2} \, e^{-2 \phi}
    \; G_{\ssM\ssP\ssQ} {G_\ssN}^{\ssP\ssQ} + \kappa^2 e^{- \phi} \;
    F_{\ssM\ssP} {F_\ssN}^\ssP
    + \frac12 \, (\Box\,
    \phi)\, g_{\ssM\ssN}  \,.\nn
\eea
In what follows we also set $G_{\ssM\ssN\ssP} = 0$.

There is an ever-growing literature on the static exact solutions
to these equations, describing compactifications of 6D down to 4D
\cite{NS,SLED1,SLED2,GGP,GGPplus,HypersNonzero}, as well as 4D de
Sitter solutions \cite{6DdSSUSY}, time-dependent solutions to the
linearized equations \cite{Linearized,KickRB,Sushastability} and
exact scaling solutions \cite{Scaling1,Scaling2}.

\subsection{Nonlinear pp-waves}

We now consider a remarkable set of exact gravitational-wave solutions to the field equations, \pref{fieldequations}. This new
class of solutions generalizes the familiar pp-wave configurations to the warped compactifications of the 6D gauged chiral
supergravity. More precisely they are 6D versions of known 4D solutions belonging to the Kundt class \cite{exactsolutionsbook},
for which pp-waves are a special case (see also \cite{warpedppsolutions} for other specific examples of these types of
solutions). There is a large literature on closely related solutions \cite{Coley}. Such pp-waves arise in a wide range of
contexts, and have been useful to understand black-hole collisions \cite{blackholecollisions}, Penrose limits of the AdS-CFT
correspondence \cite{penroselimits}, as toy models of cosmological solutions \cite{Tolley:2005us}, as well as being some of the
few known exact solutions of the string RG equations valid to all orders in $\alpha'$, i.e. consistent string backgrounds
\cite{ppwavestrings}.

For the conventional pp-wave solution the metric takes the form
\be
    \exd s^2= 2 \, \exd u \,\exd v + H(u,\vec{x}) \,\exd u^2
    + \exd \vec{x} \,^2,
\ee
where $u = x_1 + t$ and $v = x_1 - t$ are light-cone variables,
and $\vec{x}$ collectively denotes the remaining spatial
directions. The wave profile, $H$, typically satisfies an equation
of the form
\be
    \nabla^2 H = J(u),
\ee
where $J(u)$ is a source which may contain contributions from
$u$-dependent dilaton gradients and form-fields (see \cite{Blau}
for examples of relevance to string theory). Plane waves are a
special case of pp-waves for which the wave profile takes the form
$H = H_{ij}(u) x^i x^j$, representing nonlinear extensions of
gravitational perturbations around Minkowksi spacetime.

We wish to extend this class of solutions to describe
gravitational waves moving through the warped compactifications of
refs.~\cite{SLED1,SLED2,GGP,GGPplus}. To this end we make an
$\ansatz$ motivated by the solutions of ref.~\cite{GGP}, but
allowing for a more general gravitational-wave configuration
motivated by the above pp-wave solutions and their Kundt class
generalizations. For the remainder of this section we adopt
ref.~\cite{GGP}'s units, for which $\kappa^2 = \frac12$, and so
our $\ansatz$ for the metric then is
\ba
    \exd s^2 &=& W^2(\eta,u) \, e^{-\xi_1{(\eta,u)}} \Bigl[ 2\,
    \d u\, \d v + H(\eta,u,\vec{x},\theta)\, \d u^2 \Bigr] \nn\\
    &&\qquad\qquad\qquad + W^2(\eta,u) \, e^{\xi_1{(\eta,u)}}
    \Bigl[ e^{\xi_2(\eta,u)} \, \d x_1^2 + e^{-\xi_2(\eta,u)}
    \, \d x_2^2 \Bigr] \\
    && \qquad\qquad\qquad\qquad
    +K_{\eta}(\eta,u)\, \d \eta\, \d u +  a^2(\eta,u) \Bigl[
    \d \theta^2  + W^8(\eta,u) N^2(\eta,u) \, \d \eta^2
    \Bigr] \,,\nn
\ea
where the coordinates $(\eta,\theta)$ parameterize the two
internal dimensions, $(x_1,x_2)$ label the directions parallel to
the wavefront in the noncompact four dimensions and $(u,v)$ are
light-cone coordinates along the direction of wave motion. This
form is chosen so that $g_{vu}$ is the only nonzero component of
type $g_{v\ssM}$. Similarly the only non-zero component of type
$g^{u\ssM}$ is $g^{uv}$. This feature considerably simplifies the
equations of motion, and in particular implies that the inner
product, $V\cdot W$, of any two vectors, $V_\ssM$ and $W_\ssN$,
receives no contribution from terms of the form $V_{u} W_u$.
Similar properties hold for the contractions of tensors of
arbitrary rank.

As it stands this metric is more general than is necessary since
coordinate transformations of the form
\be
    v \rightarrow v + f(\eta,u) \quad \hbox{and} \quad
    \eta \rightarrow g(\eta,u) \,,
\ee
can be used to restrict some of the undetermined functions. In particular we can always set $N$ to unity by means of an
appropriate $\eta$ redefinition, and then set the resulting $K_{\eta}$ to zero by means of a redefinition of $v$. The net result
just gives a redefinition of $H(\eta,u,\vec{x},\theta)$. In what follows we shall set $K_{\eta}$ to zero, but we do not (yet) fix
$N$, for reasons which become clear below. This leads to:
\ba \label{ppwave1}
    \exd s^2 &=&  W^2(\eta,u) \, e^{-\xi_1{(\eta,u)}} \Bigl[ 2\,
    \d u\, \d v + H(\eta,u,\vec{x},\theta) \, \d u^2 \Bigr] \nn\\
    &&\qquad\qquad + W^2(\eta,u) \, e^{\xi_1{(\eta,u)}}
    \Bigl[ e^{\xi_2(\eta,u)} \, \d x_1^2 + e^{-\xi_2(\eta,u)}
    \, \d x_2^2 \Bigr] \\
    && \qquad\qquad\qquad\qquad
    +  a^2(\eta,u) \Bigl[ \d \theta^2 + W^8(\eta,u)
    \, N^2(\eta,u) \, \d \eta^2 \Bigr] \,. \nn
\ea
The fact that $H$ can depend on $\theta$ breaks the axial symmetry
corresponding to shifting this coordinate, and if $a$ is defined
so that $\theta$ goes from $0$ to $2\pi$ then it is natural to
consider periodic functions
\be
    H(\eta,u,\vec{x},\theta+2\pi)=H(\eta,u,\vec{x},\theta) \,,
\ee
to ensure regularity of the metric at finite $\eta$. We further
assume that the dilaton is taken to be a function $\phi =
\phi(\eta,u)$ and that the only nonzero component of the gauge
potential is $A_{\theta} = A_{\theta}(\eta,u)$.

The analysis of the equations of motion is greatly simplified if
we make the choice of variables \cite{GGP}
\ba
    &&\ln a = \frac14 \, (3\cx + \cy +2\cz) \nn\\
    &&\ln W = \frac14 \, (\cy-\cx) \\
    &&\phi = \frac12 \, (\cx-\cy-2\cz) \,, \nn
\ea
and so $e^{2\cx} = a^2 e^\phi$, $e^{2\cy} = a^2 W^8 e^\phi$ and
$e^{-\cz} = W^2 e^\phi$.

Substituting the \ansatz, \pref{ppwave1}, into the field
equations, \pref{fieldequations}, we find that the nontrivial
equations of motion fall into two sets: ($i$) One set of equations
--- $E$, $E_\theta$, $E_{\eta\eta}$, $E_{uv}$, $E_{11}$, $E_{22}$,
and $E_{\theta\theta}$ --- contain no $u$ derivatives, and so
degenerate into {\it ordinary} differential equations which govern
the dependence on $\eta$; ($ii$) By contrast, the second set of
equations --- $E_{uu}$ and $E_{u\eta}$ --- contain both $\eta$ and
$u$ derivatives.

We start with the first set of equations, which involve only
derivatives with respect to $\eta$. The gauge field equation
becomes
\be
    (e^{-2\cx}N^{-1} A_{\theta}')'=0,\qquad (E_{{\theta}})
\ee
where $'$ denotes $\d/\d\eta$. Convenient combinations of the
Einstein and dilaton equations then give
\ba
    (N^{-1}\xi_2')' &=&0 \qquad (E^1{}_1-E^2{}_2) \nn\\
    (N^{-1} \cz')' &=&0 \qquad (E^u{}_u +(E^1{}_1+E^2{}_2)/2) \nn\\
    (N^{-1}\xi_1')' &=&0 \qquad (E^u{}_u-(E^1{}_1+E^2{}_2)/2) \\
    N^{-1}(N^{-1}\cx')'+e^{-2\cx}\kappa^2 N^{-2}(A_{\theta}')^2 &=& 0
    \qquad (E^{\theta}{}_{\theta}) \nn\\
    N^{-1}(N^{-1}\cy')'+\frac{4g^2}{\kappa^2}e^{2\cy} &=& 0 \qquad
    (E_{\varphi}-E^{\theta}{}_{\theta}/2-E^u{}_u-(E^1{}_1+E^2{}_2)/2)
    \,. \nn
\ea
Finally the constraint equation becomes
\be \label{Constraint1}
    N^{-1} \left( \cx'^2 - \cy'^2 + \cz'^2 + {\xi_1'}^2 +
    \frac{{\xi_2'}^2}{2} + \kappa^2 e^{-2\cx} A_{\theta}'^2
    \right) - \frac{4g^2}{\kappa^2} \, e^{2\cy}
    N =0. \qquad(E^{\eta}{}_{\eta}-E^\ssM{}_\ssM /2) \,.
\ee
This is a constraint equation in the sense that it contains no
second-order derivatives with respect to $\eta$. In particular, if
it is satisfied at any given $\eta$ one can show that the other
equations imply it must be satisfied for all $\eta$.

These equations of motion, combined with the constraint, may be
derived from the following action
\be
\label{action}
    S=\int \d \eta\left[ N^{-1}\left(
    \cx'^2 - \cy'^2 + \cz'^2 + {\xi_1'}^2 + \frac{{\xi_2'}^2}{2}
    +e^{-2\cx}\kappa^2A_{\theta}'^2\right)+\frac{4
    g^2}{\kappa^2}e^{2\cy}N \right].
\ee
We now see that our reason for not gauging away the $N$ variable
was to use variation of $N$ to allow the constraint to be derived
from this action. From now on we set $N=1$ in the field equations.

Of the second set of equations, $E_{u\eta}$ is also a constraint
equation in the sense that it does not contain double $\eta$
derivatives. Its explicit form is
\be \label{Constraint2}
    2\kappa^2 e^{-2 \cx} A'_{\theta} A_{\theta,u}
    + 2\, \xi_1' \xi_{1,u}
    + \xi_2' \xi_{2,u} + 2\, \cx' \cx_{,u} - 2\, \cy' \cy_{,u}
    + 2\, \cz' \cz_{,u}
    + \xi'_{1,u} + 2\, \cy'_{,u} + \cz'_{,u} = 0 \,. \qquad
    (E_{u\eta})
\ee
Direct use of the Bianchi identities $\nabla_\ssM
G^{\ssM}{}_{\ssN}=0$ shows that
\be
    \partial_{\eta} E_{u\eta} = c_{AB}(\eta,u)
    E^{AB} + d(\eta,u) \partial_u E_{uv} \,.
\ee
Thus provided the other equations are satisfied, and we choose the
boundary data so that $E_{u\eta}=0$ on a surface $\eta=\eta_0$,
then this equation is satisfied for all $\eta$.

The final equation $(E_{uu})$ is a linear, sourced wave equation
for the wave profile $H$. It takes the form
\be
\label{sourceequation}
    \hat{O} H = J(\eta,u) \qquad (E_{uu}) \,,
\ee
where $\hat{O}$ is proportional to the scalar Laplacian on the
spacetime (\ref{ppwave1}),
\be
    \hat{O}=\partial^2_\eta + e^{-\xi_1-\xi_2+2\cy+\cz} \,
    \partial^2_1 + e^{-\xi_1+\xi_2+2\cy+\cz}\, \partial^2_2
    + e^{-2\cx+2\cy} \, \partial^2_\theta \,,
\ee
and the source, $J$, is given by
\ba
\label{source}
    J(\eta,u) &=& -e^{\xi_1 + 2\cy+\cz} \Bigl[ 2\kappa^2
    e^{-2\cx} A_{\theta,u}^2
    + 3\, \xi_{1,u}^2 + \xi_{2,u}^2 + 2\, \cx_{,u}^2 + 2\, \cy_{,u}^2
    + 4\cy_{,u} \cz_{,u} \nn\\
    &&\qquad\qquad  + 3\, \cz_{,u}^2 +2 \, \xi_{1,u}
    (2 \, \cy_{,u} + \cz_{,u}) + 2\, \xi_{1,uu}
    + 4\,\cy_{,uu} + 2\,\cz_{,uu}
    \Bigr] \,.
\ea
This form is reminiscent of the usual pp-wave equations, except
that now the source additionally depends on the direction $\eta$.

The whole system of equations is now straightforward to solve. We
first solve the ordinary differential equations for
$(A_\theta, \cx, \cy, \cz, \xi_1, \xi_2)$, and promote the
integration constants to arbitrary functions of $u$. Choosing $N =
1$, this leads to
\ba
\label{explict1}
    &&A_\theta' = C_1(u) \,  e^{2\cx} \,, \quad
    \cz = C_2(u) \, \eta + C_3(u) \nn\\
    && \xi_1 = C_4(u) \, \eta + C_5(u) \,, \quad
    \xi_2 = C_6(u) \, \eta + C_7(u) \,,
\ea
and
\ba
    &&\cx'' + \kappa^2  C^2_1(u) \, e^{2\cx} = 0 \nn\\
    &&\cy'' + \frac{4\, g^2}{\kappa^2} \, e^{2\cy} = 0 \,.
\ea
The explicit solutions to these equations are
\ba
\label{explicit2}
&&  e^{\cx} =  \frac{C_8(u)}{\kappa \, C_1(u)} {\rm sech} \(C_8(u) (\eta-\eta_1(u)) \)\, , \nn \\
&&  e^{\cy} =   \kappa \, \frac{C_9(u)}{2g} {\rm sech}\(C_9(u) (\eta-\eta_2(u)) \)        \, .
\ea
The restrictions on the integration `constants',
$C_i(u)$, appearing here come from imposing the constraints
--- {\it i.e.} eqs.~\pref{Constraint1} and \pref{Constraint2} ---
on some chosen surface $\eta=\eta_0$. In addition we have the flux quantization constraint which comes from
defining the gauge field to be smooth on two separate patches and demanding that the associated gauge transformation picks up an integer multiple of $2\pi$ phase
on integration around $\theta$. In practise this is the statement that
\be
\int \int  F_{\eta \, \theta} \, \d \theta \d \eta = \frac{2\pi n}{q},
\ee
where $q$ is the smallest unit of fermion charge coupled to $A_{\mu}$. This is equivalent to
\be
\int_{-\infty}^{\infty} C_1(u) \, e^{2\cx} \d \eta = \frac{n}{q}.
\ee
Substituting in the above solution Eq.~(\ref{explicit2}) for $\cx$, this relation fixes the function $C_1(u)$ to be
\be
C_1(u)=\frac{2 q \, C_8(u)}{\kappa^2n}.
\ee
Finally we can compute the source $J(\eta,u)$ to obtain the wave
profile $H$ in the form
\be
    H(\eta,u,x_1,x_2,\theta)=\int_{\eta_0}^{\eta} \d \eta_1
    \int_{\eta_0}^{\eta_1} \d\eta_2 \; J(\eta_2,u)
    + H_0(\eta,u,x_1,x_2,\theta) \,,
\ee
where $H_0(\eta,u,x_1,x_2,\theta)$ is any solution of the
homogeneous equation: $\hat{O} H_0 =0$. Since $\hat{O}$ does not
contain $u$ derivatives, in solving this equation we may further
allow all the integration constants to be arbitrary functions of
$u$. As is the case for normal pp-waves, this leaves us free to
introduce an arbitrary functional dependence on $u$ into the wave
profile.

\subsection{Breaking the null symmetry}

A simple interesting extension of the wave solutions can be
obtained by allowing for the wave profile $H$ to depend on the
null coordinate $v$. Once this is done there is no longer a null
Killing vector. The $\ansatz$ that is consistent with the
equations of motion in this case is
\ba
    \exd s^2 &=& W^2(\eta,u) \, e^{-\xi_1{(\eta,u)}} \Bigl[ 2\, \d
    u\,\d v + H(\eta,u,\vec{x},\theta,v) \,\d u^2 \Bigr] \nn\\
    &&\qquad\qquad
    +  W^2(\eta,u) \, e^{\xi_1{(\eta,u)}} \Bigl[ e^{\xi_2(\eta,u)} \, \d x_1^2
    + e^{-\xi_2(\eta,u)} \, \d x_2^2 \Bigr] \\
    &&\qquad\qquad\qquad\qquad + a^2(\eta,u) \Bigl[ \d \theta^2
    + W^8(\eta,u) \d \eta^2 \Bigr] \,, \nn
\ea
with
\be
    H(\eta,u,\vec{x},\theta,v) = H_1(\eta,u,\vec{x},\theta)
    + H_2(\eta,u) v + H_3(u) v^2 \,.
\ee
This is identical to the situation for the 4D solutions belonging
to the Kundt class \cite{exactsolutionsbook}. From now on we shall focus on the case for which $H_3(u)$ is independent of $u$.
The analysis proceeds as before. In particular, the set of equations involving only derivatives with respect to
$\eta$ can be encoded in the action
\ba
    S &=& \int \d \eta \left[ N^{-1} \left( \cx'^2 - \cy'^2
    + \cz'^2 + {\xi_1'}^2 + \frac{{\xi_2'}^2}{2}
    + \kappa^2 e^{-2\cx} {A_{\theta}'}^2
    \right) \right.\nn\\
    && \qquad\qquad\qquad\qquad\qquad\qquad \left.
    + \left( \frac{4 g^2}{\kappa^2} \, e^{2\cy} - 2 \, H_3
    \, e^{\xi_1+2\cy+\cz} \right) N \right],
\ea
while the $E_{\eta u}$ constraint becomes
\be
    H_2' = 2\kappa^2 e^{-2\cx} A'_{\theta}A_{\theta,u}
    + 2\, \xi_1' \xi_{1,u} + \xi_2' \xi_{2,u} + 2\, \cx'
    \cx_{,u} - 2\, \cy' \cy_{,u} + 2\, \cz' \cz_{,u} + \xi'_{1,u}
    + 2 \,\cy'_{,u} + \cz'_{,u} \,, \;\;
    (E_{+\eta})
\ee
which we regard as an equation to be solved for $H_2$.
Substituting this back into the $E_{uu}$ equation we find
\be
    \hat{O} H_1 =
    \tilde{J}, \qquad (E_{++})
\ee
with the new source given by
\ba
    \tilde{J} &=& - e^{\xi_1 + 2\cy + \cz} \left(2 \kappa^2 e^{-2\cx}
    A_{\theta,u}^2 + 3\, \xi_{1,u}^2 + \xi_{2,u}^2 + 2\, \cx_{,u}^2
    +2\, \cy_{,u}^2 + 4\, \cy_{,u} \cz_{,u}
    + 3\, \cz_{,u}^2 \right. \nn \\
    && \left. + 2\, \xi_{1,u} (2\, \cy_{,u} + \cz_{,u})
    + 2\,\xi_{1,uu}
    +4\, \cy_{,uu} + 2 \,\cz_{,uu} \right)
    - \partial_u(\xi_1 + 2 \, \cy +
    \cz)H_2 \,.
\ea
These equations are again straightforward to solve (in principle).

\subsection{Relation to brane properties}

An important property of the static compactifications is the
connection between their asymptotic forms near their singularities
and the physical properties of the branes present at these
singularities \cite{GGPplus,NavSant,6DdSSUSY,Scaling1,UVcaps}. These show
that in the case where gravity is weak the near-brane limit of the
extrinsic curvatures are related to the components of the
effective 4D brane stress energy.

\subsubsection*{Special case: pure tension branes}

In particular, bulk solutions which are sourced by pure-tension branes (in general the tension may also be dilaton dependent
$T=T(\phi)$) must have asymptotic limits for which the spatial and temporal parts of the metric in the noncompact four dimensions
are invariant under 4D Lorentz transformations. We next identify the conditions which the above solutions must satisfy in order
to be sourced in this way by this type of Lorentz-invariant brane.

In this case the symmetry of the boundary conditions requires
$\xi_1' = 0$, $\xi_2' = 0$ and $H' = 0$ at the positions of each
brane. Inspection of the solutions shows that the first two of
these conditions require $\xi_1' = \xi_2' =0$ everywhere
throughout the bulk, and so we may for convenience everywhere set
$\xi_1 = \xi_2 = 0$. On so doing, the remaining equations for
$\cx$, $\cy$, $\cz$ and $A_{\theta}$ are identical to those
obtained for the static solutions of ref.~\cite{GGP,GGPplus}, and
we find then the most general solution to have the form:
\bea \label{GGPvsu}
    e^{-\phi(\eta,u)} &=& W^{2} e^{\lambda_3(u) \eta} \nn\\
    W^4(\eta,u) &=& \left( \frac{Q (u) \lambda_2(u)}{4g\lambda_1(u)}
    \right)
    \frac{\cosh[ \lambda_1(u)(\eta - \eta_1(u))]}{\cosh[ \lambda_2(u)
    (\eta - \eta_2(u))]} \\
    a^{-4}(\eta,u) &=& \left( \frac{gQ^3(u)}{\lambda_1^3(u)
    \lambda_2(u)}
    \right) e^{-2\lambda_3(u)\eta} \cosh^3[\lambda_1(u)(\eta
    - \eta_1(u))]
    \cosh[ \lambda_2(\eta - \eta_2(u))] \nonumber\\
    A^{\pm}_{\theta}(\eta,u) &=& F_{\pm} + \int^{\eta}_{\pm \infty} \d \eta \left(
    \frac{Q(u)
    a^2 \, e^{-\lambda_3(u)\eta}}{W^2}  \right) \,.\nn
\eea
These are precisely the static solutions, but with all integration constants made into functions of $u$, for which the branes are
situated at $\eta = \pm \infty$. The choice of $A^{\pm}_{\theta}$ corresponds to the the fact that we strictly need two patches to define a smooth gauge field. As discussed earlier, one function of $u$ (e.g. $Q(u)$) will be fixed by the flux quantization condition which is here the statement that
\be
\int^{\infty}_{-\infty} \d \eta \left(\frac{Q(u)a^2 \, e^{-\lambda_3(u)\eta}}{W^2}  \right)=\frac{n}{q}
\ee
and so
\be
Q(u)=\frac{4q\lambda_1(u)}{n}.
\ee
The $E_{\eta\eta}$ constraint imposes the relation $\lambda_2^2(u)
= \lambda_3^2(u) + \lambda_1^2(u)$. By choosing the boundary data
for the $\eta$ integration at the surface $\eta = \eta_0
\rightarrow -\infty$ we see that the $E_{\eta u}$ constraint
becomes at leading order
\be
    \lambda_2\lambda_{2,u} = \lambda_3\lambda_{3,u}
    + \lambda_1\lambda_{1,u} \,,
\ee
which is just the $u$ derivative of the $E_{\eta\eta}$ constraint,
and is hence automatically satisfied. At subleading order we find $\(Q(u)/\lambda_1(u)\)_{,u}=0$ which is automatically satisfied if the flux quantization condition is satisfied. In other words this constraint does not give us any new information on the integration functions.

The last constraint required for pure-tension branes is the
imposition of $H' = 0$ at $\eta = \pm \infty$. In general we see
no obstruction to achieving this using the above solution. For
instance, suppose that $H$ were only to depend on $\eta$ and $u$.
Then we may use the homogenous solution to arrange $H' \to 0$ at
one brane. This is consistent with $H' = 0$ at the other brane if
the quantity $I(u) = 0$, where
\ba
    I(u) &\equiv& \int_{-\infty}^{+\infty} \d \eta \, J(\eta,u)  \\
    &=& - \int_{-\infty}^{+\infty} \d \eta \, e^{2\cy+\cz}
    \left(2\kappa^2 e^{-2\cx}{A_{\theta,u}}^2 + 2\, \cx_{,u}^2
    +2\, \cy_{,u}^2 + 4\, \cy_{,u} \cz_{,u} + 3\cz_{,u}^2
    +4\, \cy_{,uu} + 2\, \cz_{,uu} \right) \,. \nn
\ea
We must first check to see if this integral is convergent. In general convergence of all but the first term in brackets is
guaranteed by the exponential falloff of $e^{2\cy+\cz}$ at both boundaries. On the other hand, we also require ${A_{\theta,u}}^2$
to falls of at the branes as $e^{2\cx}$ and this is sufficient to guarantee the convergence of the integral as a whole. Given
that the integral converges, and that the $u$-dependence of the integration constants is arbitrary, it would seem to be possible
in general to choose this $u$-dependence to arrange $I(u) = 0$, and so $H'=0$.

\section{Applications}

The solutions we described in the previous section represent the
natural lift of the 4 dimensional pp-waves and Kundt class
solutions up to 6 dimensions, and consequently they have many of
the same applications as arise for usual pp-waves.

\subsection{Shock waves arising from high energy particles on the
brane}

A classic example of a pp-wave is the Aichelberg-Sexl metric
\cite{Aichelburg:1970dh} which describes the metric induced by an
ultra-relativistic particle moving through Minkowski spacetime,
\be
    \exd s^2 = 2 \exd u \, \exd v - \frac{\kappa^2 m}{\pi} \delta(u)
    \log(|\vec{x}|) \exd u^2 + \exd\vec{x}^{\,2}.
\ee
It is straightforward to obtain this metric by taking the
Schwarzshild metric, which describes the gravitational field of a
massive particle at rest, and infinitely boosting it (i.e. taking
the Penrose limit) so that the particle is moving at the speed of
light. Alternatively we may obtain the metric directly as the
pp-wave induced by a stress-energy of the form
\be
    T_{uu} = m \, \delta(u) \delta^2(\vec{x}) \,,
\ee
which is localised on the worldline of a massless particle
with energy $m$. This metric is useful in understanding how black
holes form in the collision of two high energy particles
\cite{blackholecollisions}.

In the present context it is natural to ask, what is the 6D metric
describing a high energy particle traveling  on one of the
branes? Not surprisingly the solution is given by one of the
warped pp-waves we described in the previous section, provided we supplement the equations with an additional source term
\be
T_{uu}=m \, \delta(u) \, \delta^2(\vec{x}) \, \delta^2(\vec{y}-\vec{y}_b),
\ee
where $\vec{y}_b$ denotes the position of one of the branes, and as before $m$ is the energy of the ultra-relativistic particle on the brane. The only equation that is modified from before is $E_{uu}$, so that (\ref{sourceequation}) is replaced with
\be
    \hat{O} H = J(\eta,u) -2 \kappa^2 m \, \delta(u) \, \delta^2(\vec{x}) \, \delta^2(\vec{y}-\vec{y}_b).
\ee
A crucial point is that because this source is traceless (both in a six dimensional $T^M_M=0$ and four dimensional sense $T^{\mu}_{\mu}=0$), the positions of the branes are not affected by the presence of the source. Thus the entire backreaction of the source is encoded in the modified wave profile for $H$.

Rather than finding the most general solution consistent with this boundary condition, we shall look for solutions which have an $O(2)$ symmetry under rotations in the $\vec{x}$ directions, which amounts to choosing $\xi_2=0$, hence $C_6(u)=C_7(u)=0$ and are in addition $\theta$ independent so that $\partial_{\theta} H=0$. Both of these are consistent truncations given the form of the source. To reiterate the form of the bulk metric is
\ba
    \exd s^2 &=&  W^2(\eta,u) \, e^{-\xi_1{(\eta,u)}} \Bigl[ 2\,
    \d u\, \d v + H(\eta,u,\vec{x},\theta) \, \d u^2 \Bigr] \nn\\
    &&+ W^2(\eta,u) \, e^{\xi_1{(\eta,u)}}\d^2{\vec{x}}
    +  a^2(\eta,u) \Bigl[ \d \theta^2 + W^8(\eta,u)
    \, \d \eta^2 \Bigr] \,, \nn
\ea
where
\be
    A_\theta' = C_1(u) \,  e^{2\cx} \,, \quad
    \cz = C_2(u) \, \eta + C_3(u), \, \quad
     \xi_1 = C_4(u) \, \eta + C_5(u) ,
\ee
and
\ba
&&  e^{\cx} =  \frac{C_8(u)}{\kappa \, C_1(u)} {\rm sech} \(C_8(u) (\eta-\eta_1(u)) \)\, , \nn \\
&&  e^{\cy} =   \kappa \, \frac{C_9(u)}{2g} {\rm sech}\(C_9(u) (\eta-\eta_2(u)) \)        \, ,
\ea
with $C_1(u)=2\, q \, C_8(u) \slash \kappa^2 n$ from flux quantization. We must also check that the constraints are satisfied. Equation (\ref{Constraint1}) amounts to
\be
C_9^2(u)=C_8^2(u)+C_2^2(u)+C_4^2(u),
\ee
whereas on substituting this relation into equation (\ref{Constraint2}) we get
\be
\(\frac{C_1(u)}{C_8(u)} \)_{,u}=0.
\ee
which is consistent with the flux quantization condition and so does not give us a new constraint.
The exact nonlinear equation for the wave profile $H$ reduces to
\be
\left[\partial_{\eta}^2+e^{-\xi_1+2\cy+\cz} \vec{\nabla}^2 \right]H=J(\eta,u)-2\kappa^2 m \delta(u) \frac{\delta(r)}{2\pi r}\frac{W_b^4\delta(\eta-\eta_b)}{2\pi},
\ee
where we write $r^2=\vec{x}{}^2$. Because the two branes are located at $\eta \rightarrow \pm \infty$ in our coordinates, we adopt the artifice of taking the position of the brane to be at $\eta=\eta_b$ with $\eta_b \rightarrow \pm \infty$, depending on which brane the particle is on, taken at the end of the calculation.
The expression for the source $J(\eta,u)$, which is now fully determined, is given by (\ref{source}). Let us consider the behavior of the source as $\eta \rightarrow \pm \infty$. It is easy to see that the leading behaviour is
\be
J \sim \eta^2 e^{(C_4(u)+C_2(u)) \eta-2|C_9(u)| |\eta|},
\ee
and so provided $|C_9(u)| > |C_4(u)+C_2(u)|/2$, $J$ is bounded at $\eta = \pm \infty$. Given this we can integrate the profile as
\be
H(\eta,u,\vec{x})=\int^{\eta}_{-\infty} \d \eta_1 \int^{\eta_1}_{-\infty} \d \eta_2 \, J(\eta_2,u) +H_0(\eta,u,\vec{x}),
\ee
where $H_0$ satisfies
\be
\left[\partial_{\eta}^2+e^{-\xi_1+2\cy+\cz} \vec{\nabla}^2 \right]H_0=-2\kappa^2 m \delta(u) \frac{\delta(r)}{2\pi r}\frac{W_b^4\delta(\eta-\eta_b)}{2\pi}.
\ee
The formal solution to this equation can be expressed as the shock wave
\be
H_0(\eta,r,u)=-2\kappa^2 m \, \delta(u) \int_0^{\infty} \frac{\d k}{2\pi} k \, J_0(kr)\, u_k(\eta,u),
\ee
where the modes $u_k(\eta,u)$ are defined to satisfy
\be
    \left[ \partial_{\eta}^2 - \frac{\kappa^2C_9(u)^2}{4g^2}\frac{e^{(C_2(u)-C_4(u))\eta+C_3(u)-C_5(u)}}{{\rm cosh}
    ^2\(C_9(u) (\eta-\eta_2(u))\)} \, k^2\right] u_k(\eta,u) = W_b^4 \frac{\delta (\eta-\eta_{b})}{2\pi}
    \,.
\ee
To see how to proceed let us choose to place the source on the brane at $\eta_b =-\infty$. We then look for a solution to the homogeneous equation which is regular at $\eta=+\infty$ and satisfies the boundary condition
\be
\frac{\partial}{\partial \eta}u_k(\eta,u)|_{\eta=\eta_b}=\frac{W_b^4}{2\pi}.
\ee
In principal this is straightforward to do numerically. As a specific analytical example let us choose $C_4(u)=C_2(u)$.
The general solutions of the homogenous equations can be obtained in terms of the variable $z={\rm tanh}(C_9(u) (\eta-\eta_2(u)))$.
It is given in terms of Legendre polynomials:
\be
u_k(\eta,u)= A_k P_{\nu}(z)+B_k Q_{\nu}(z) ,
\ee
with
\be
\nu=\frac{1}{2}\left[-1+\sqrt{1-\frac{k^2\kappa^2 C_9^2(u) }{g^2}\, e^{C_3(u)-C_5(u)}} \right].
\ee
We can without loss of generality choose $C_9(u) > 0 $ so that the sourced brane is at $z=-1$. Demanding regularity at $z=+1$ enforces $B_k=0$. Note that this is still the correct solution even when $\nu$ is complex, and that $P_{\nu}(z)$ is still real. The Legendre function has the property that
\be
\lim_{z \rightarrow -1} P_{\nu}(z) = -\frac{2\gamma+{\ln}(\frac{1+z}{2})+\psi_{-\nu}(0)+\psi_{1+\nu}(0)}{\Gamma_{-\nu}\Gamma_{1+\nu}} +\dots \, ,
\ee
with $\psi$ the polygamma function. As is usual for codimension two objects, these metrics strictly diverge logarithmically at the branes, seen here as the logarithmic divergence of the Legendre polynomial at $z=-1$. This issue may be understood in terms of the regularization
and renormalization prescription described in Ref.~\cite{claudiaeft}. For now we shall just regularize this by hand. Finally, from the boundary condition we can determined $A_k$
\be
A_k=-  2C_9(u)\Gamma_{-\nu}\Gamma_{1+\nu} \frac{W_b^4}{2\pi} \, .
\ee
This completes the specification of the solution.

These metrics could be useful in understanding black hole
collisions when the impact parameter is comparable to the size of
the extra dimensions, for which it is necessary to take properly
into account the warping in the extra dimensions.

\subsection{Brane Bubble Walls}

Another physical application of the above solutions is to the problem of
understanding how 6D compactifications respond to phase
transitions on the source branes. It is known that such a phase
transition can significantly perturb the bulk field
configurations, and in some circumstances might completely
destabilize it. This expectation comes because it is known that
static solutions can only exist provided that the tensions and
other bulk couplings of the two source branes are related to one
another \cite{SLED1,GP,GGPplus}. More recently it has been found
that the late-time solutions which arise when the branes are not
so adjusted are time-dependent scaling solutions \cite{Scaling1,Scaling2}.
This suggests that if the tension on one brane were to be
perturbed in an arbitrary way, it could precipitate a transition
from a static background towards one of the runaway scaling
solutions.

One way to estimate this response is to perform a linearized
stability analysis about one of the known static compactifications
\cite{Linearized,KickRB,Sushastability}. The response to a change in one of the
brane tensions may be accounted for in such a calculation by
allowing the perturbations to become singular at the brane
positions in a known way \cite{KickRB}. The result of such a
calculation is that the perturbations are marginally stable,
allowing time-dependent deformations to grow along the system's
known flat direction.

\begin{figure}
    \begin{center}
     \includegraphics[width=0.6\textwidth]{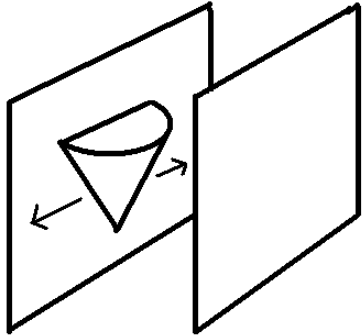}
    \end{center}
     \caption{A cartoon of a wave moving into the bulk and along the brane due to a bubble-nucleation event on a brane. The equations
in the text provide an approximate description far from the nucleation site, and near the brane (where the disturbance is
approximately a plane wave moving parallel to a brane direction).}
   \label{BraneWave}
\end{figure}

This type of linearized calculation is not really
satisfactory because it models the brane tension as changing
simultaneously throughout the brane volume. However causality
implies that such a transition must occur locally with the
nucleation of a small region having the new tension at some
position on the brane, with the region of new tension then
expanding to fill space to take more and more of the brane volume
to the new vacuum. The resulting geometry is likely to be
reasonably complicated, with a wave propagating out into the bulk
as well as along the brane directions (see Fig.~\ref{BraneWave}).
We expect the scaling solutions of ref.~\cite{Scaling1,Scaling2} to be
obtained after the wave reflects back and forth a sufficient
number of times around the internal dimensions.

The solutions in this paper provide some approximate insight into
the nature of the waves which would carry the news of the phase
transition across the brane and the bulk. The insight is only
approximate partly because the wave solutions we consider are
plane waves moving along a direction parallel to the branes,
rather than describing `spherical' emission into the brane and the
bulk from the nucleation event. Our solution might nevertheless
capture some of the physics of the outgoing wave far from the
nucleation site, along directions in spacetime near to the brane
itself (and so moving close to parallel to a brane direction).
Since it describes a single wave it also can only apply before any
waves have had time to be reflected back to the initial brane
after crossing the extra dimensions. In such a situation we expect
the physics of wave propagation to be largely insensitive to the
specific boundary conditions taken by our solutions on the
spectator brane. These conditions need not be inconsistent with
one another, particularly in scenarios for which the extra
dimensions are much larger than the particle-physics scales which
govern the nucleation event itself.

The change in brane properties with wave passage can then be read off from the $u$-dependence of the integration `constants' in
the pp-wave solutions. For instance, for the pure-tension branes the quantities $\lambda_i(u)$ appearing in eqs.~\pref{GGPvsu}
can be related to the time-dependent brane stress-energies by formulae such as those given in refs.~\cite{NavSant,Scaling1}, with
the result that the 4D tension of the brane located at $\eta \to \pm \infty$ is proportional to the quantity $3\,\alpha_\pm +
\beta_\pm$, where
\be \label{GGPPowers}
    \alpha_\pm(u) = \frac{\lambda_2(u) - \lambda_1(u)}{5
    \lambda_2(u) - \lambda_1(u)
    \mp 2 \lambda_3(u)}  \quad \hbox{and} \quad
    \beta_\pm(u) = \frac{\lambda_2(u) + 3 \lambda_1(u) \mp 2\lambda_3(u)}{5
    \lambda_2(u) - \lambda_1(u) \mp 2 \lambda_3(u)} \,.
\ee

\section{Conclusions}

Six dimensional supergravity provides a fruitful laboratory for
investigating the issues which underly higher-dimensional physics
in general, and brane approaches to the cosmological constant
problem in particular. It does so because 6D is rich enough to
exhibit many of the properties of still-higher dimensions --- like
moduli-stabilization through fluxes \cite{Susha}, brane
back-reaction on internal geometries, chiral fermions and
Green-Schwarz anomaly cancellation \cite{6Danomalies}. Yet it is
also on the one hand simple enough to allow the development of
techniques for obtaining physically interesting exact solutions,
but on the other hand not so simple as to be misleading about what
happens in higher dimensional (in a way which co-dimension one
physics sometimes can be).

We have used these properties to explore solution-generation
techniques which we believe to be applicable to a wide variety of
higher-dimensional supergravities. We do so by using these
techniques to construct a new class of time-dependent exact
solutions to the field equations of 6D chiral gauged supergravity.
These solutions describe the physics of nonlinear gravitational
waves passing through compactified spacetimes for which two
dimensions are self-consistently compactified in response to the
presence of two space-filling branes and a bulk Maxwell flux.

We believe these methods to merit more detailed exploration.

\section*{Acknowledgements}

The work of A.J.T. and C.d.R. at the
Perimeter Institute is supported in part by the Government of Canada through NSERC and by the Province of Ontario through MRI.
CB would like to thank the Banff International Research Station
for its kind hospitality while part of this work was being completed. CB and DH are supported in part by funds from
Natural Sciences and Engineering Research Council of Canada, and
CB also acknowledges the Killam Foundation and McMaster University
for research support.


\begin{thebibliography}{99}

\bibitem{HiDSugra}
For a survey of many of the higher-dimensional supergravities see,
for example, {\it Supergravities in Diverse Dimensions} Vols. I \&
II, ed. by A. Salam and E. Sezgin, World Scientific 1989.

\bibitem{tdepsusy}

%
 M.~Gutperle and A.~Strominger,
  ``Spacelike Branes,''
  JHEP {\bf 0204} (2002) 018
  [hep-th/0202210];
  %%CITATION = HEP-TH 0202210;%%
%
 C.~P.~Burgess, F.~Quevedo, S.~J.~Rey, G.~Tasinato and I.~Zavala,
  ``Cosmological spacetimes from negative tension brane backgrounds,''
  JHEP {\bf 0210} (2002) 028
  [hep-th/0207104];
  %%CITATION = HEP-TH 0207104;%%
%
 N.~Ohta,
  ``Accelerating cosmologies from S-branes,''
  Phys.\ Rev.\ Lett.\  {\bf 91} (2003) 061303
  [hep-th/0303238];
  %%CITATION = HEP-TH 0303238;%%
%
  C.~P.~Burgess, C.~Nunez, F.~Quevedo, G.~Tasinato and I.~Zavala,
   ``General brane geometries from scalar potentials: Gauged supergravities  and
  accelerating universes,''
  JHEP {\bf 0308} (2003) 056
  [hep-th/0305211];
  %%CITATION = HEP-TH 0305211;%%
  %
    C.~M.~Chen, P.~M.~Ho, I.~P.~Neupane, N.~Ohta and J.~E.~Wang,
  ``Hyperbolic Space Cosmologies,''
  JHEP {\bf 0310} (2003) 058
  [hep-th/0306291];
  %%CITATION = HEP-TH 0306291;%%
  %
   L.~Cornalba and M.~S.~Costa,
  ``Time-dependent orbifolds and string cosmology,''
  Fortsch.\ Phys.\  {\bf 52} (2004) 145
  [hep-th/0310099];
  %%CITATION = HEP-TH 0310099;%%
  %
    M.~Cariglia, G.~W.~Gibbons, R.~Guven and C.~N.~Pope,
  ``Non-Abelian pp-waves in D = 4 supergravity theories,''
  Class.\ Quant.\ Grav.\  {\bf 21} (2004) 2849
  [hep-th/0312256];
  %%CITATION = HEP-TH 0312256;%%
  %
  G.~Tasinato, I.~Zavala, C.~P.~Burgess and F.~Quevedo,
  ``Regular S-brane backgrounds,''
  JHEP {\bf 0404} (2004) 038
  [hep-th/0403156];
  %%CITATION = HEP-TH 0403156;%%
  %
   P.~Chen, K.~Dasgupta, K.~Narayan, M.~Shmakova and M.~Zagermann,
  ``Brane inflation, solitons and cosmological solutions: I,''
  JHEP {\bf 0509} (2005) 009
  [hep-th/0501185];
  %%CITATION = HEP-TH 0501185;%%
  %
   E.~A.~Bergshoeff, A.~Collinucci, D.~Roest, J.~G.~Russo and P.~K.~Townsend,
  ``Cosmological D-instantons and cyclic universes,''
  Class.\ Quant.\ Grav.\  {\bf 22} (2005) 2635
  [hep-th/0504011];
  %%CITATION = HEP-TH 0504011;%%
  %
   M.~Cvetic, G.~W.~Gibbons, H.~Lu and C.~N.~Pope,
   ``Rotating Black Holes In Gauged Supergravities: Thermodynamics,
  Supersymmetric Limits, Topological Solitons And Time Machines,''
  [hep-th/0504080];
  %%CITATION = HEP-TH 0504080;%%
%\cite{Hellerman:2006hf}
%\bibitem{Hellerman:2006hf}
  S.~Hellerman and I.~Swanson,
  ``Cosmological unification of string theories,''
  arXiv:hep-th/0612116.
  %%CITATION = HEP-TH/0612116;%%
%\cite{Hellerman:2006nx}
%\bibitem{Hellerman:2006nx}
  S.~Hellerman and I.~Swanson,
  ``Cosmological solutions of supercritical string theory,''
  arXiv:hep-th/0611317.
  %%CITATION = HEP-TH/0611317;%%

\bibitem{tdepnonsusy}

 P.~K.~Townsend and M.~N.~R.~Wohlfarth,
  ``Accelerating cosmologies from compactification,''
  Phys.\ Rev.\ Lett.\  {\bf 91} (2003) 061302
  [hep-th/0303097];
  %%CITATION = HEP-TH 0303097;%%
  %
    S.~B.~Giddings and R.~C.~Myers,
  ``Spontaneous decompactification,''
  Phys.\ Rev.\ D {\bf 70} (2004) 046005
  [hep-th/0404220];
  %%CITATION = HEP-TH 0404220;%%
  %
   G.~W.~Gibbons, H.~Lu, D.~N.~Page and C.~N.~Pope,
  ``Rotating black holes in higher dimensions with
  a cosmological constant,''
  Phys.\ Rev.\ Lett.\  {\bf 93} (2004) 171102
  [hep-th/0409155].
  %%CITATION = HEP-TH 0409155;%%

\bibitem{RS}
  L.~Randall and R.~Sundrum,
  ``A large mass hierarchy from a small extra dimension,''
  Phys.\ Rev.\ Lett.\  {\bf 83} (1999) 3370
  [hep-ph/9905221];
  %%CITATION = HEP-PH 9905221;%%
  %
  %  L.~Randall and R.~Sundrum,
  ``An alternative to compactification,''
  Phys.\ Rev.\ Lett.\  {\bf 83} (1999) 4690
  [hep-th/9906064].
  %%CITATION = HEP-TH 9906064;%%

\bibitem{SLED1}
Y. Aghababaie, C.P. Burgess, S. Parameswaran and F. Quevedo,
Nucl.\ Phys.\ {\bf B680} (2004) 389--414, [hep-th/0304256];
%%CITATION = HEP-TH 0304256;%%
%
%\bibitem{TAMU}
  C.~P.~Burgess,
  ``Towards a natural theory of dark energy: Supersymmetric large extra
  dimensions,''
  AIP Conf.\ Proc.\  {\bf 743} (2005) 417
  [hep-th/0411140].
  %%CITATION = HEP-TH 0411140;%%

\bibitem{SLED2}
%
%\cite{Aghababaie:2003ar}
%\bibitem{Aghababaie:2003ar}
  Y.~Aghababaie {\it et al.},
  ``Warped brane worlds in six dimensional supergravity,''
  JHEP {\bf 0309}, 037 (2003)
  [arXiv:hep-th/0308064].
  %%CITATION = JHEPA,0309,037;%%

\bibitem{SLEDx}
%
C.P. Burgess, ``Supersymmetric Large Extra Dimensions and the
Cosmological Constant: An Update,'' {\it Ann. Phys.} {\bf 313}
(2004) 283-401 [hep-th/0402200];
%%CITATION = HEP-TH 0402200;%%
%
  J.~Garriga and M.~Porrati,
  ``Football shaped extra dimensions and the absence of self-tuning,''
  JHEP {\bf 0408} (2004) 028
  [hep-th/0406158].
  %%CITATION = HEP-TH 0406158;%%

\bibitem{UVSensitivity}
 C.~P.~Burgess and D.~Hoover,
  ``UV sensitivity in supersymmetric large extra dimensions: The Ricci-flat
  case,''
  [hep-th/0504004];
  %%CITATION = HEP-TH 0504004;%%
  %
   D.~M.~Ghilencea, D.~Hoover, C.~P.~Burgess and F.~Quevedo,
  ``Casimir energies for 6D supergravities compactified on T(2)/Z(N) with
  Wilson lines,''
  JHEP {\bf 0509}, 050 (2005)
  [hep-th/0506164];
  %%CITATION = HEP-TH 0506164;%%
  %
  D.~Hoover and C.~P.~Burgess,
  ``Ultraviolet sensitivity in higher dimensions,''
  JHEP {\bf 0601}, 058 (2006)
  [hep-th/0507293].
  %%CITATION = HEP-TH 0507293;%%

\bibitem{SLEDpheno}
  G.~Azuelos, P.~H.~Beauchemin and C.~P.~Burgess,
  ``Phenomenological constraints on extra-dimensional scalars,''
  J.\ Phys.\ G {\bf 31}, 1 (2005)
  [hep-ph/0401125];
  %%CITATION = HEP-PH 0401125;%%
  %
  C.~P.~Burgess, J.~Matias and F.~Quevedo,
  ``MSLED: A minimal supersymmetric large extra dimensions scenario,''
  Nucl.\ Phys.\ B {\bf 706} (2005) 71
  [hep-ph/0404135];
  %%CITATION = HEP-PH 0404135;%%
  %
   P.~H.~Beauchemin, G.~Azuelos and C.~P.~Burgess,
  ``Dimensionless coupling of bulk scalars at the LHC,''
  J.\ Phys.\ G {\bf 30}, N17 (2004)
  [hep-ph/0407196];
  %%CITATION = HEP-PH 0407196;%%
  %
    J.~Matias and C.~P.~Burgess,
  ``MSLED, neutrino oscillations and the cosmological constant,''
  JHEP {\bf 0509} (2005) 052
  [hep-ph/0508156];
  %%CITATION = HEP-PH 0508156;%%
  %
   P.~Callin and C.~P.~Burgess,
  ``Deviations from Newton's law in supersymmetric large extra dimensions,''
  [hep-ph/0511216].
  %%CITATION = HEP-PH 0511216;%%



\bibitem{5DSelfTune}
N.~Arkani-Hamed, S.~Dimopoulos, N.~Kaloper and R.~Sundrum, ``A
small cosmological constant from a large extra dimension,'' Phys.\
Lett.\ B {\bf 480} (2000) 193, [hep-th/0001197];
%
S.~Kachru, M.~B.~Schulz and E.~Silverstein, ``Self-tuning flat
domain walls in 5d gravity and string theory,'' Phys.\ Rev.\ D
{\bf 62} (2000) 045021, [hep-th/0001206].

\bibitem{5DSelfTunex}
S.~Forste, Z.~Lalak, S.~Lavignac and H.~P.~Nilles, ``A comment on
self-tuning and vanishing cosmological constant in the  brane
world'', Phys.\ Lett.\ B {\bf 481} (2000) 360, hep-th/0002164;
JHEP {\bf 0009} (2000) 034, [hep-th/0006139];\\
%
C.~Csaki, J.~Erlich, C.~Grojean and T.J.~Hollowood, ``General
Properties of the Self-Tuning Domain Wall Approach to the
Cosmological Constant Problem,'' Nucl.\ Phys.\ {\bf B584} (2000)
359-386, [hep-th/0004133];\\
%
C.~Csaki, J.~Erlich and C.~Grojean, ``Gravitational Lorentz
Violations and Adjustment of the Cosmological Constant in
Asymmetrically Warped Spacetimes,'' Nucl.\ Phys.\ {\bf B604}
(2001) 312-342, [hep-th/0012143];\\
%
J.M. Cline and H. Firouzjahi, ``No-Go Theorem for Horizon-Shielded
Self-Tuning Singularities'', Phys.\ Rev.\ {\bf D65} (2002) 043501,
[hep-th/0107198].

\bibitem{6DNonSUSYSelfTune}
J.-W. Chen, M.A. Luty and E. Pont{\'o}n, JHEP 0009 (2000) 012,
[hep-th/0003067];
%
S.~M.~Carroll and M.~M.~Guica, ``Sidestepping the cosmological
constant with football-shaped extra dimensions,''
[hep-th/0302067];
%
I.~Navarro, ``Co-dimension two compactifications and the
cosmological constant  problem,'' JCAP {\bf 0309} (2003) 004
[hep-th/0302129].
%%CITATION = HEP-TH 0302129;%%
%
%\cite{Schwindt:2005fm}
%\bibitem{Schwindt:2005fm}
  J.~M.~Schwindt and C.~Wetterich,
  ``Dark energy cosmologies for codimension-two branes,''
  Nucl.\ Phys.\  B {\bf 726}, 75 (2005)
  [arXiv:hep-th/0501049].
  %%CITATION = NUPHA,B726,75;%%
%
  %\cite{Schwindt:2005ns}
%\bibitem{Schwindt:2005ns}
  J.~M.~Schwindt and C.~Wetterich,
  ``The cosmological constant problem in codimension-two brane models,''
  Phys.\ Lett.\  B {\bf 628}, 189 (2005)
  [arXiv:hep-th/0508065].
  %%CITATION = PHLTA,B628,189;%%


\bibitem{6DNonSUSYSelfTunex}
I.~Navarro, ``Spheres, deficit angles and the cosmological
constant,'' Class.\ Quant.\ Grav.\  {\bf 20} (2003) 3603
[hep-th/0305014];
%%CITATION = HEP-TH 0305014;%%
%
H.~P.~Nilles, A.~Papazoglou and G.~Tasinato, ``Selftuning and its
footprints,'' Nucl.\ Phys.\ B {\bf 677} (2004) 405
[hep-th/0309042];
%%CITATION = HEP-TH 0309042;%%
%
  P.~Bostock, R.~Gregory, I.~Navarro and J.~Santiago,
  ``Einstein gravity on the Co-dimension 2 brane?,''
  Phys.\ Rev.\ Lett.\  {\bf 92}, 221601 (2004)
  [hep-th/0311074];
  %%CITATION = HEP-TH 0311074;%%
%
  J.~Vinet and J.~M.~Cline,
  ``Can Co-dimension-two branes solve the cosmological constant problem?,''
  Phys.\ Rev.\ D {\bf 70} (2004) 083514
  [hep-th/0406141];
  %%CITATION = HEP-TH 0406141;%%
  %
  M.~L.~Graesser, J.~E.~Kile and P.~Wang,
  ``Gravitational perturbations of a six dimensional self-tuning model,''
  Phys.\ Rev.\ D {\bf 70} (2004) 024008
  [hep-th/0403074];
  %%CITATION = HEP-TH 0403074;%%
%
  %
  G.~Kofinas,
  ``On braneworld cosmologies from six dimensions, and absence thereof,''
  [hep-th/0506035].
  %%CITATION = HEP-TH 0506035;%%

\bibitem{Burgess:2007ui}
  C.~P.~Burgess,
  ``Extra Dimensions and the Cosmological Constant Problem,''
  arXiv:0708.0911 [hep-ph].
  %%CITATION = ARXIV:0708.0911;%%

\bibitem{Closelyrelatedworks}

%\cite{Salvio:2006mh}
%\bibitem{Salvio:2006mh}
  A.~Salvio,
  ``4D effective theory and geometrical approach,''
  AIP Conf.\ Proc.\  {\bf 881}, 58 (2007)
  [arXiv:hep-th/0609050].
  %%CITATION = APCPC,881,58;%%
  %
  %\cite{Parameswaran:2006db}
%\bibitem{Parameswaran:2006db}
  S.~L.~Parameswaran, S.~Randjbar-Daemi and A.~Salvio,
  %``Gauge fields, fermions and mass gaps in 6D brane worlds,''
  Nucl.\ Phys.\  B {\bf 767}, 54 (2007)
  [arXiv:hep-th/0608074].
  %%CITATION = NUPHA,B767,54;%%
  %
  %\cite{Salvio:2007mb}
%\bibitem{Salvio:2007mb}
  A.~Salvio,
  %``Aspects of physics with two extra dimensions,''
  arXiv:hep-th/0701020.
  %%CITATION = HEP-TH/0701020;%%
  %
  %\cite{Peloso:2006cq}
%\bibitem{Peloso:2006cq}
  M.~Peloso, L.~Sorbo and G.~Tasinato,
  %``Standard 4d gravity on a brane in six dimensional flux
  %compactifications,''
  Phys.\ Rev.\  D {\bf 73}, 104025 (2006)
  [arXiv:hep-th/0603026].
  %%CITATION = PHRVA,D73,104025;%%
 %
%\cite{Kobayashi:2007qe}
%\bibitem{Kobayashi:2007qe}
  T.~Kobayashi and Y.~i.~Takamizu,
  ``Hybrid compactifications and brane gravity in six dimensions,''
  arXiv:0707.0894 [hep-th].
  %%CITATION = ARXIV:0707.0894;%%
%
%\cite{Gogberashvili:2007gg}
%\bibitem{Gogberashvili:2007gg}
  M.~Gogberashvili, P.~Midodashvili and D.~Singleton,
  ``Fermion Generations from `Apple-Shaped' Extra Dimensions,''
  arXiv:0706.0676 [hep-th].
  %%CITATION = ARXIV:0706.0676;%%
%
%\cite{Lee:2007ib}
%\bibitem{Lee:2007ib}
  H.~M.~Lee and A.~Papazoglou,
  ``Gravitino in six-dimensional warped supergravity,''
  arXiv:0705.1157 [hep-th].
  %%CITATION = ARXIV:0705.1157;%%
%\cite{Elizalde:2007di}
%\bibitem{Elizalde:2007di}
  E.~Elizalde, M.~Minamitsuji and W.~Naylor,
  ``Casimir effect in rugby-ball type flux compactifications,''
  Phys.\ Rev.\  D {\bf 75}, 064032 (2007)
  [arXiv:hep-th/0702098].
  %%CITATION = PHRVA,D75,064032;%%
M.~Cvetic, G.~W.~Gibbons and C.~N.~Pope,
``A string and M-theory origin for the Salam-Sezgin model,''
  Nucl. Phys.  B 677, 164 (2004)
  [arXiv:hep-th/0308026].


\bibitem{regularizations}

%\cite{Geroch:1987qn}
%\bibitem{Geroch:1987qn}
  R.~Geroch and J.~H.~Traschen,
  ``Strings and Other Distributional Sources in General Relativity,''
  Phys.\ Rev.\  D {\bf 36}, 1017 (1987).
  %%CITATION = PHRVA,D36,1017;%%
  %
  %\cite{Cline:2003ak}
%\bibitem{Cline:2003ak}
  J.~M.~Cline, J.~Descheneau, M.~Giovannini and J.~Vinet,
  ``Cosmology of codimension-two braneworlds,''
  JHEP {\bf 0306}, 048 (2003)
  [arXiv:hep-th/0304147].
  %%CITATION = JHEPA,0306,048;%%
%
%\cite{de Rham:2005ci}
%\bibitem{de Rham:2005ci}
  C.~de Rham and A.~J.~Tolley,
  ``Gravitational waves in a codimension two braneworld,''
  JCAP {\bf 0602}, 003 (2006)
  [arXiv:hep-th/0511138].
  %%CITATION = JCAPA,0602,003;%%
%
%\cite{Fujii:2007fi}
%\bibitem{Fujii:2007fi}
  S.~Fujii, T.~Kobayashi and T.~Shiromizu,
  ``Low energy effective theory on a regularized brane in six-dimensional flux
  compactifications,''
  arXiv:0708.2534 [hep-th].
  %%CITATION = ARXIV:0708.2534;%%
  %
  %\cite{Papantonopoulos:2006dv}
%\bibitem{Papantonopoulos:2006dv}
  E.~Papantonopoulos, A.~Papazoglou and V.~Zamarias,
  ``Regularization of conical singularities in warped six-dimensional
  compactifications,''
  JHEP {\bf 0703}, 002 (2007)
  [arXiv:hep-th/0611311].
  %%CITATION = JHEPA,0703,002;%%
  %
  %\cite{Papantonopoulos:2007fk}
%\bibitem{Papantonopoulos:2007fk}
  E.~Papantonopoulos, A.~Papazoglou and V.~Zamarias,
  ``Induced cosmology on a regularized brane in six-dimensional flux
  compactification,''
  arXiv:0707.1396 [hep-th].
  %%CITATION = ARXIV:0707.1396;%%
%
  %
  %\cite{Minamitsuji:2007fx}
%\bibitem{Minamitsuji:2007fx}
  M.~Minamitsuji and D.~Langlois,
  ``Cosmological evolution of regularized branes in 6D warped flux
  compactifications,''
  arXiv:0707.1426 [hep-th].
  %%CITATION = ARXIV:0707.1426;%%
%
%\cite{Himmetoglu:2006nw}
%\bibitem{Himmetoglu:2006nw}
  B.~Himmetoglu and M.~Peloso,
  ``Isolated Minkowski vacua, and stability analysis for an extended brane in
  the rugby ball,''
  Nucl.\ Phys.\  B {\bf 773}, 84 (2007)
  [arXiv:hep-th/0612140].
  %%CITATION = NUPHA,B773,84;%%

\bibitem{claudiaeft}
W.~D.~Goldberger and M.~B.~Wise,
  %``Renormalization group flows for brane couplings,''
  Phys.\ Rev.\  D {\bf 65} (2002) 025011
  [arXiv:hep-th/0104170].
  %%CITATION = PHRVA,D65,025011;%%
  %
  %\cite{de Rham:2007pz}
%\bibitem{de Rham:2007pz}
  C.~de Rham,
  ``The Effective Field Theory of Codimension-two Branes,''
  JHEP {\bf 0801}, 060 (2008)
  [arXiv:0707.0884 [hep-th]].
  %%CITATION = JHEPA,0801,060;%%
%\bibitem{deRham:2007dg}
  C.~de Rham,
  ``Classical Renormalization of Codimension-two Brane Couplings,''
  AIP Conf.\ Proc.\  {\bf 957}, 309 (2007)
  [arXiv:0710.4598 [hep-th]].
  %%CITATION = APCPC,957,309;%%



\bibitem{NS}
H. Nishino and E. Sezgin, {\it Phys. Lett.} {\bf 144B} (1984) 187;
``The Complete N=2, D = 6 Supergravity With Matter And Yang-Mills
Couplings,'' Nucl.\ Phys.\ {\bf B278} (1986) 353;
%%CITATION = NUPHA,B278,353;%%
%
S. Randjbar-Daemi, A. Salam, E. Sezgin and J. Strathdee, {\it
Phys. Lett.} {\bf B151} (1985) 351; A.~Salam and E.~Sezgin,
``Chiral Compactification On Minkowski $\times S^2$ Of N=2
Einstein-Maxwell Supergravity In Six-Dimensions,'' Phys.\ Lett.\ B
{\bf 147} (1984) 47.
%%CITATION = PHLTA,B147,47;%%

\bibitem{6DSugra}
Other 6D supergravities are discussed in
%
  N. Marcus and J.H. Schwarz, Phys.\ Lett.\ B {\bf 115} (1982)
  111;
%
  R.~D'Auria, P.~Fre and T.~Regge,
  ``Consistent Supergravity In Six-Dimensions Without Action Invariance,''
  Phys.\ Lett.\ B {\bf 128} (1983) 44;
  %%CITATION = PHLTA,B128,44;%%
%
  Y.~Tanii,
  ``N=8 Supergravity In Six-Dimensions,''
  Phys.\ Lett.\ B {\bf 145} (1984) 197;
  %%CITATION = PHLTA,B145,197;%%
%
L.J. Romans, Nucl.\ Phys.\ {\bf B269} (1986) 691--711.

\bibitem{GandC}
S.~Weinberg, {\sl Gravitation and Cosmology}, Wiley, New York,
1972.

\bibitem{MTW}
C.W.~Misner, K.P.~Thorne and J.A.~Wheeler, {\sl Gravitation}, W.H.
Freeman and Company (1970).

\bibitem{GGP}
  G.~W.~Gibbons, R.~Guven and C.~N.~Pope,
  ``3-branes and uniqueness of the Salam-Sezgin vacuum,''
  Phys.\ Lett.\ B {\bf 595}, 498 (2004)
  [hep-th/0307238].
  %%CITATION = HEP-TH 0307238;%%

\bibitem{GGPplus}
  C.~P.~Burgess, F.~Quevedo, G.~Tasinato and I.~Zavala,
  ``General axisymmetric solutions and self-tuning in 6D chiral gauged
  supergravity,''
  JHEP {\bf 0411}, 069 (2004)
  [hep-th/0408109].
  %%CITATION = HEP-TH 0408109;%%

\bibitem{HypersNonzero}
   S.~Randjbar-Daemi and E.~Sezgin,
  ``Scalar potential and dyonic strings in 6d gauged supergravity,''
  Nucl.\ Phys.\ B {\bf 692} (2004) 346
  [hep-th/0402217];
  %%CITATION = HEP-TH 0402217;%%
%
 A.~Kehagias,
  ``A conical tear drop as a vacuum-energy drain for the solution of the
  cosmological constant problem,''
  Phys.\ Lett.\ B {\bf 600} (2004) 133
  [hep-th/0406025];
  %%CITATION = HEP-TH 0406025;%%
%
  S.~Randjbar-Daemi and V.~A.~Rubakov,
  ``4d-flat compactifications with brane vorticities,''
  JHEP {\bf 0410}, 054 (2004)
  [hep-th/0407176];
  %%CITATION = HEP-TH 0407176;%%
%
  H.~M.~Lee and A.~Papazoglou,
  ``Brane solutions of a spherical sigma model in six dimensions,''
  Nucl.\ Phys.\ B {\bf 705} (2005) 152
  [hep-th/0407208];
  %%CITATION = HEP-TH 0407208;%%
%
 V.~P.~Nair and S.~Randjbar-Daemi,
  ``Nonsingular 4d-flat branes in six-dimensional supergravities,''
  JHEP {\bf 0503} (2005) 049
  [hep-th/0408063];
  %%CITATION = HEP-TH 0408063;%%
%
   S.~L.~Parameswaran, G.~Tasinato and I.~Zavala,
  ``The 6D SuperSwirl,''
  [hep-th/0509061];
  %%CITATION = HEP-TH 0509061;%%
%
 H.~M.~Lee and C.~Ludeling,
  ``The general warped solution with conical branes in six-dimensional
  supergravity,''
  [hep-th/0510026].
  %%CITATION = HEP-TH 0510026;%%

\bibitem{6DdSSUSY}
  A.~Tolley, C.P.~Burgess, D.~Hoover and Y.~Aghababaie,
  ``Bulk Singularities and the Effective Cosmological Constant
   for Higher Co-dimension Branes,''
   JHEP {\bf 0603} (2006) 091 [hep-th/0512218].

\bibitem{Linearized}
%\cite{Cline:2003ak}
%\bibitem{Cline:2003ak}
  J.~M.~Cline, J.~Descheneau, M.~Giovannini and J.~Vinet,
  ``Cosmology of codimension-two braneworlds,''
  JHEP {\bf 0306}, 048 (2003)
  [arXiv:hep-th/0304147].
  %%CITATION = JHEPA,0306,048;%%
 %
  H.~M.~Lee and A.~Papazoglou,
  ``Scalar mode analysis of the warped Salam-Sezgin model,''
  [hep-th/0602208];
  %%CITATION = HEP-TH 0602208;%%

\bibitem{KickRB}
  %\cite{Burgess:2006ds}
%\bibitem{Burgess:2006ds}
  C.~P.~Burgess, C.~de Rham, D.~Hoover, D.~Mason and A.~J.~Tolley,
  ``Kicking the rugby ball: Perturbations of 6D gauged chiral supergravity,''
  JCAP {\bf 0702}, 009 (2007)
  [arXiv:hep-th/0610078].
  %%CITATION = JCAPA,0702,009;%%

\bibitem{Sushastability}
%\cite{Parameswaran:2007cb}
%\bibitem{Parameswaran:2007cb}
  S.~L.~Parameswaran, S.~Randjbar-Daemi and A.~Salvio,
  ``Stability and Negative Tensions in 6D Brane Worlds,''
  arXiv:0706.1893 [hep-th].
  %%CITATION = ARXIV:0706.1893;%%


\bibitem{Scaling1}
%\cite{Tolley:2006ht}
%\bibitem{Tolley:2006ht}
  A.~J.~Tolley, C.~P.~Burgess, C.~de Rham and D.~Hoover,
  ``Scaling solutions to 6D gauged chiral supergravity,''
  New J.\ Phys.\  {\bf 8}, 324 (2006)
  [arXiv:hep-th/0608083].
  %%CITATION = NJOPF,8,324;%%

\bibitem{Scaling2}

%\cite{Copeland:2007ur}
%\bibitem{Copeland:2007ur}
  E.~J.~Copeland and O.~Seto,
  ``Dynamical solutions of warped six dimensional supergravity,''
  arXiv:0705.4169 [hep-th].
  %%CITATION = ARXIV:0705.4169;%%
%\cite{Kobayashi:2007hf}
%\bibitem{Kobayashi:2007hf}
%
  T.~Kobayashi and M.~Minamitsuji,
  ``Brane cosmological solutions in six-dimensional warped flux
  compactifications,''
  arXiv:0705.3500 [hep-th].
  %%CITATION = ARXIV:0705.3500;%%
%\cite{Burgess:2007vi}

\bibitem{exactsolutionsbook}
%\cite{Stephani:2003tm}
%\bibitem{Stephani:2003tm}
  H.~Stephani, D.~Kramer, M.~MacCallum, C.~Hoenselaers and E.~Herlt,
  ``Exact solutions of Einstein's field equations,''
%\href{http://www.slac.stanford.edu/spires/find/hep/www?irn=5562600}{SPIRES entry}
{\it  Cambridge, UK: Univ. Pr. (2003) 701 P}

%\cite{Kaloper:2006ek}
\bibitem{warpedppsolutions}
  N.~Kaloper and D.~Kiley,
  ``Exact black holes and gravitational shockwaves on
  codimension-2 branes,''
  JHEP {\bf 0603}, 077 (2006)
  [arXiv:hep-th/0601110].
  %%CITATION = JHEPA,0603,077;%%
%
%\cite{Anber:2007ry}
%\bibitem{Anber:2007ry}
  M.~Anber and L.~Sorbo,
  ``Two gravitational shock waves on the AdS3 brane,''
  arXiv:0706.1560 [hep-th].
  %%CITATION = ARXIV:0706.1560;%%
  %
  %\cite{Kaloper:2005wa}
%\bibitem{Kaloper:2005wa}
  N.~Kaloper,
  ``Gravitational shock waves and their scattering
  in brane-induced  gravity,''
  Phys.\ Rev.\  D {\bf 71}, 086003 (2005)
  [Erratum-ibid.\  D {\bf 71}, 129905 (2005)]
  [arXiv:hep-th/0502035].
  %%CITATION = PHRVA,D71,086003;%%
  %
  %\cite{Kaloper:2005az}
%\bibitem{Kaloper:2005az}
  N.~Kaloper,
  ``Brane-induced gravity's shocks,''
  Phys.\ Rev.\ Lett.\  {\bf 94}, 181601 (2005)
  [Erratum-ibid.\  {\bf 95}, 059901 (2005)]
  [arXiv:hep-th/0501028].
  %%CITATION = PRLTA,94,181601;%%
  %
  D.~Kiley,
  ``Rotating Black Holes on Codimension-2 Branes,''
  arXiv:0708.1016 [hep-th].
  %%CITATION = ARXIV:0708.1016;%%
%
%\cite{Burgess:2007ui}


\bibitem{Coley}

%\cite{Coley:2007tp}
%\bibitem{Coley:2007tp}
  A.~A.~Coley,
  ``Classification of the Weyl Tensor in Higher Dimensions and Applications,''
  arXiv:0710.1598 [gr-qc].
  %%CITATION = ARXIV:0710.1598;%%
%
%\cite{Coley:2007rk}
%\bibitem{Coley:2007rk}
  A.~Coley, A.~Fuster and S.~Hervik,
  ``Supergravity solutions with constant scalar invariants,''
  arXiv:0707.0957 [hep-th].
  %%CITATION = ARXIV:0707.0957;%%
%
%\cite{Coley:2007yx}
%\bibitem{Coley:2007yx}
  A.~A.~Coley, A.~Fuster, S.~Hervik and N.~Pelavas,
  ``Vanishing Scalar Invariant Spacetimes in Supergravity,''
  JHEP {\bf 0705}, 032 (2007)
  [arXiv:hep-th/0703256].
  %%CITATION = JHEPA,0705,032;%%
%
%\cite{Coley:2006fr}
%\bibitem{Coley:2006fr}
  A.~Coley, A.~Fuster, S.~Hervik and N.~Pelavas,
  ``Higher dimensional VSI spacetimes,''
  Class.\ Quant.\ Grav.\  {\bf 23}, 7431 (2006)
  [arXiv:gr-qc/0611019].
  %%CITATION = CQGRD,23,7431;%%
%
%\cite{Coley:2004hu}
%\bibitem{Coley:2004hu}
  A.~Coley, R.~Milson, V.~Pravda and A.~Pravdova,
  ``Vanishing scalar invariant spacetimes in higher dimensions,''
  Class.\ Quant.\ Grav.\  {\bf 21}, 5519 (2004)
  [arXiv:gr-qc/0410070].
  %%CITATION = CQGRD,21,5519;%%
%
%\cite{Coley:2004yc}
%\bibitem{Coley:2004yc}
  A.~A.~Coley and S.~Hervik,
  ``Brane waves,''
  Class.\ Quant.\ Grav.\  {\bf 21}, 5759 (2004)
  [arXiv:gr-qc/0405089].
  %%CITATION = CQGRD,21,5759;%%
%
%\cite{Coley:2004jv}
%\bibitem{Coley:2004jv}
  A.~Coley, R.~Milson, V.~Pravda and A.~Pravdova,
  ``Classification of the Weyl tensor in higher-dimensions,''
  Class.\ Quant.\ Grav.\  {\bf 21}, L35 (2004)
  [arXiv:gr-qc/0401008].
  %%CITATION = CQGRD,21,L35;%%


\bibitem{blackholecollisions}

%\cite{Griffiths:1991zp}
%\bibitem{Griffiths:1991zp}
  J.~B.~Griffiths,
  ``Colliding plane waves in general relativity,''
%\href{http://www.slac.stanford.edu/spires/find/hep/www?irn=2459426}{SPIRES entry}
{\it  Oxford, UK: Clarendon (1991) 232 p. (Oxford mathematical
monographs).}
%
%\cite{Eardley:2002re}
%\bibitem{Eardley:2002re}
  D.~M.~Eardley and S.~B.~Giddings,
  ``Classical black hole production in high-energy collisions,''
  Phys.\ Rev.\  D {\bf 66}, 044011 (2002)
  [arXiv:gr-qc/0201034].
  %%CITATION = PHRVA,D66,044011;%%
%\cite{Kaloper:2007pb}
%\bibitem{Kaloper:2007pb}
  N.~Kaloper and J.~Terning,
  ``How black holes form in high energy collisions,''
  Gen.\ Rel.\ Grav.\  {\bf 39}, 1525 (2007)
  [arXiv:0705.0408 [hep-th]].
  %%CITATION = GRGVA,39,1525;%%

\bibitem{penroselimits}

%\cite{Maldacena:2003nj}
%\bibitem{Maldacena:2003nj}
  J.~M.~Maldacena,
  ``Lectures on AdS/CFT,''
  arXiv:hep-th/0309246.
  %%CITATION = HEP-TH/0309246;%%
%
%\cite{Pankiewicz:2003pg}
%\bibitem{Pankiewicz:2003pg}
  A.~Pankiewicz,
  ``Strings in plane wave backgrounds,''
  Fortsch.\ Phys.\  {\bf 51}, 1139 (2003)
  [arXiv:hep-th/0307027].
  %%CITATION = FPYKA,51,1139;%%

%\cite{Tolley:2005us}
\bibitem{Tolley:2005us}
  A.~J.~Tolley,
  ``String propagation through a big crunch / big bang transition,''
  Phys.\ Rev.\  D {\bf 73}, 123522 (2006)
  [arXiv:hep-th/0505158].
  %%CITATION = PHRVA,D73,123522;%%

\bibitem{ppwavestrings}

%\cite{Horowitz:1990xq}
%\bibitem{Horowitz:1990sr}
  G.~T.~Horowitz and A.~R.~Steif,
  ``Strings in strong gravitational fields,''
  Phys.\ Rev.\  D {\bf 42}, 1950 (1990).
  %%CITATION = PHRVA,D42,1950;%%
%\cite{Horowitz:1989bv}
%\bibitem{Horowitz:1989bv}
  G.~T.~Horowitz and A.~R.~Steif,
  ``Space-Time Singularities in String Theory,''
  Phys.\ Rev.\ Lett.\  {\bf 64}, 260 (1990).
  %%CITATION = PRLTA,64,260;%%

\bibitem{Blau}
%\cite{Blau:2002js}
  M.~Blau and M.~O'Loughlin,
  ``Homogeneous plane waves,''
  Nucl.\ Phys.\ B {\bf 654}, 135 (2003)
  [arXiv:hep-th/0212135].
  %%CITATION = HEP-TH 0212135;%%
  M.~Blau, M.~O'Loughlin, G.~Papadopoulos and A.~A.~Tseytlin,
  ``Solvable models of strings in homogeneous plane wave backgrounds,''
  Nucl.\ Phys.\ B {\bf 673}, 57 (2003)
  [arXiv:hep-th/0304198].
  %%CITATION = HEP-TH 0304198;%%

\bibitem{NavSant}
  I.~Navarro and J.~Santiago,
  ``Gravity on Co-dimension 2 brane worlds,''
  JHEP {\bf 0502}, 007 (2005)
  [hep-th/0411250].
  %%CITATION = HEP-TH 0411250;%%

\bibitem{GP}
G.~W.~Gibbons and C.~N.~Pope,
   ``Consistent S**2 Pauli reduction of six-dimensional chiral gauged
  Einstein-Maxwell supergravity,''
  Nucl.\ Phys.\ B {\bf 697} (2004) 225
  [hep-th/0307052].
  %%CITATION = HEP-TH 0307052;%%

\bibitem{Susha}
  Y.~Aghababaie, C.~P.~Burgess, S.~L.~Parameswaran and F.~Quevedo,
   ``Susy Breaking and Moduli Stabilization from Fluxes in Gauged
   6D
  Supergravity,''
  JHEP {\bf 0303} (2003) 032
  [hep-th/0212091].
  %%CITATION = HEP-TH 0212091;%%

\bibitem{6Danomalies}
M.B. Green, J.H. Schwarz and P.C. West, {\it Nucl. Phys.} {\bf
B254} (1985) 327;\\
%
J. Erler, {\it J. Math. Phys.} {\bf 35} (1994) 1819
[hep-th/9304104].


\bibitem{UVcaps}

%\cite{Burgess:2007vi}
%\bibitem{Burgess:2007vi}
  C.~P.~Burgess, D.~Hoover and G.~Tasinato,
  ``UV Caps and Modulus Stabilization for 6D Gauged Chiral Supergravity,''
  JHEP {\bf 0709}, 124 (2007)
  [arXiv:0705.3212 [hep-th]].
  %%CITATION = JHEPA,0709,124;%%

%\cite{Aichelburg:1970dh}
\bibitem{Aichelburg:1970dh}
  P.~C.~Aichelburg and R.~U.~Sexl,
  ``On the Gravitational field of a massless particle,''
  Gen.\ Rel.\ Grav.\  {\bf 2} (1971) 303.
  %%CITATION = GRGVA,2,303;%%










\end{thebibliography}
\end{document}